\documentclass{llncs}
\usepackage{graphicx}
\usepackage{amsfonts}
\usepackage{makeidx}  
\newcommand{\be}{\begin{equation}}
\newcommand{\ee}{\end{equation}}
\newcommand{\noi}{\noindent}
\newcommand{\vecvar}[1]{{\bf #1\/}}
\newcommand{\px}{\vecvar{x}}
\newcommand{\pr}{\vecvar{r}}
\newcommand{\pk}{\vecvar{k}}
\newcommand{\ignorar}[1]{}
\newcommand{\Real}{\mathbb{R}}

\newcommand{\MSC}{{\cal F}_{\infty}}
\newcommand{\msc}{{\cal F}_{_{_{\infty}}}\!\!\!\!}

\newcommand{\essgrad}[1]{\nabla_{_{_{_{\!\! #1}}}}\!\! s}

\begin{document}
\frontmatter          
\pagestyle{headings}  
%
\mainmatter              
\title{Microcanonical processing methodology for ECG and intracardial potential: application to atrial fibrillation}
\titlerunning{$\mu$canonical processing for cardiac potential}  
%
\author{Oriol Pont\inst{1} \and Michel Ha\"{\i}ssaguerre\inst{2} \and Hussein Yahia\inst{1} \and Nicolas Derval\inst{2} \and M\'el\`eze Hocini\inst{2}}
\authorrunning{Oriol Pont et al.} 
%
%
\institute{INRIA Bordeaux Sud-Ouest, Campus Bordeaux 1, 33405 Talence, France,\\
\email{oriol.pont@inria.fr},\\ WWW home page: \texttt{http://geostat.bordeaux.inria.fr/}
\and
H\^opital Cardiologique du Haut-L\'ev\^eque, Ave. de Magellan, 33604 Pessac, France}

\maketitle              

\begin{abstract}
Cardiac diseases are the principal cause of human morbidity and mortality in the western world. The electric potential of the heart is a highly complex signal emerging as a result of nontrivial flow conduction, hierarchical structuring and multiple regulation mechanisms. Its proper accurate analysis becomes of crucial importance in order to detect and treat arrhythmias or other abnormal dynamics that could lead to life-threatening conditions.
To achieve this, advanced nonlinear processing methods are needed: one example here is the case of recent advances in the Microcanonical Multiscale Formalism. The aim of the present paper is to recapitulate those advances and extend the analyses performed, specially looking at the case of atrial fibrillation. We show that both ECG and intracardial potential signals can be described in a model-free way as a fast dynamics combined with a slow dynamics. Sharp differences in the key parameters of the fast dynamics appear in different regimes of transition between atrial fibrillation and healthy cases. Therefore, this type of analysis could be used for automated early warning, also in the treatment of atrial fibrillation particularly to guide radiofrequency ablation procedures.

\end{abstract}

\section{Introduction}
Heartbeat signals emerge as the result of a nontrivial and highly complex synchronization process between the cardiac pacemaker cells which are hierarchically connected. This structure ensures the robustness of heartbeats and under normal conditions it exhibits a regular main rate perturbed by small chaotic fluctuations. Typically, the amplitude of such fluctuations is much smaller than the average interbeat interval, something that makes the healthy (sinus rhythm) heartbeat appear as mainly periodic. Nevertheless, the fluctuations around this main period are not an unstructured random noise but follow a complex dynamics. Moreover, extensive research findings in the last decade have shown that, the characterization of these fluctuations can be crucial for determining whether the heart is healthy or it is indicating signs of a transition to an arrhythmia, despite still expressing normally \cite{Ivanov.1999,Ivanov.2003,Kozaitis.2008,Martinelli.2010,Pont.2011b}.

The cardiac action potential is leaded by polarization of pacemaker cells. These cells are not homogeneously distributed but mainly concentrate in nodes (sinoatrial and atrioventricular nodes), bundles and the Purkinje fibers which 
innervate all the ventricular myocardium. The action of pacemaker cells controls the cardiac contractions (atrial and ventricular systole) and relaxation (diastole) in an organized way to ensure the optimal pumping \cite{Klabunde.2005}. Polarization and depolarization of the membrane potential is a collective effect and so they are affected by the conduction dynamics of the electric flow \cite{Khan.2009}. In this sense, the orientation of cardiomyocites \cite{Rothaus.2007} plays an important role and other structural particularities such as that of epicardial fat \cite{Coppini.2009,Coppini.2011} can have an important effect in particular cases. The complex mechanical structure of the heart together with the conductivity inhomogeneities and the hierarchical activation are themselves governed through nontrivial regulation mechanisms, which include the sympathetic and parasympathetic divisions of the autonomic nervous system and the endocrine system.

All these elements and, most importantly, their mutual interrelations make the cardiac dynamics highly complex and as a consequence its description remains mainly phenomenological and the microscopic state of the components is impossible to determine noninvasively in most cases. That is why effective descriptions provide an alternative for the automated analysis of heartbeat signals. In that context, first studies of interbeat fluctuations found that they have a multiscale structure \cite{Kitney.1980,Kobayashi.1982} and so fractal models were proposed for it \cite{Peng.1993}. Later on, the development of more advanced analysis techniques based on multiresolution analysis and characterization of singularities, large deviations and predictability permitted a more extensive study \cite{Ivanov.1999}, which shows that a healthy heartbeat has a multifractal structure, while a heart under congestive heart failure deviates from multifractality. In that study, the multifractal analysis performed is based solely on the Legendre spectrum. As a consequence, from a thermodynamic point of view, it corresponds to a canonical formalism, not microcanonical, and so local singularity exponents for each point cannot be directly obtained within the formalism. For that reason, recent methods developed under the framework of the Microcanonical Multiscale Formalism (MMF) \cite{Pont.2006,Pont.2009a,Turiel.2008} make them an especially appropriate alternative for heartbeat analysis; e.g. in \cite{Pont.2011b,Pont.2011c}.

The process resulting in the multiscale structure observed in heartbeat consists of the synchronization process in a hierarchic complex network \cite{Ribeiro.2000}; the connectivity network of cardiac pacemaker cells in our case. As a consequence, an analysis based on the singularity exponents \cite{Mallat.1991,Mallat.1992} and the optimal wavelet \cite{Pont.2009c,Pont.2011} applied to heartbeat time series allows directly accessing the geometric features that characterize their multiscale behavior. The results obtained in \cite{Ivanov.1999} are based on a canonical analysis, meaning that the behavior of statistical averages is used to indirectly retrieve the geometric features: scaling exponents of partition functions estimate a curve that can be used to obtain the so-called singularity spectrum by means of a numerically-estimated Legendre transform). This methodology is known to give less accurate estimation on the tails of the singularity spectrum for which a microcanonical analysis has been found to be much more robust and accurate \cite{Turiel.2006}. Having such estimation has a capital importance for anticipating as much as possible when heartbeat dynamics starts drifting from the healthy behavior. Given the quickness with which heart failure can be fatal or leave irreversible sequelae, the precise estimation provided by the MMF has a strong potential in helping to save lives and improve the health of people with cardiac diseases. Multifractal models originated from the study of turbulent flows. While blood can be in turbulent regime -and notably inside the heart- it is unlikely that this turbulence is reflected to the cardiac electric flow. This electric flow, and in consequence the measured heartbeat signal, are the result of a complex synchronization process between pacemakers. As explained in \cite{Chainais.2005}, traffic in a complex network under certain regimes becomes multifractal and so the analysis of the pacemaker network could explain the observed multifractality.

This paper is presented as an extended version of the preliminary results by the same signing authors that were presented in the 6th International Conference MDA 2011 in New York \cite{Pont.2011b}. Here we summarize our previous findings, we expand the technical description of the methods, we include the presented results and we extend them with a deeper analysis, a more exhaustive data processing and statistical testing.

The paper is structured as follows: in the next Section we present the methods used in our analysis. We introduce the basics of the MMF and the algorithms to accurately retrieve the empirical singularity exponents from a signal. We also present a reconstruction method and how it can be used to separate the fast dynamics implied by the singularity exponents from a slow dynamics that indicates changes in regime. The fast dynamics is a simple orientation transition without memory. The slow dynamics modulates the fast one and it can be easily sifted from it. In Section~\ref{sec:data} we introduce the empirical data to be analyzed as well as some basics of the atrial fibrillation condition that are relevant for its signal processing. In Section~\ref{sec:results} we apply that analysis to the heartbeat data and discuss how it can be used to identify dynamical changes, specially for the case of atrial fibrillation. Finally, in Section~\ref{sec:conclusions} we draw the conclusions of our work.

\section{Methods: The Microcanonical Multiscale Formalism}
\label{sec:MMF}

The MMF is a theoretical and methodological framework for the analysis of multiscale signals. Its basic element of description is by means of the singularity exponents of a signal under analysis, which are the exponents describing the local regular/singular behavior of the signal around each point \cite{Pont.2006,Pont.2009a,Turiel.2008}.

\subsection{Singularity exponents}
\label{sec:singan}

Singularity exponents have different mathematical definitions depending on the context they are used. The usual notion in complex-signal analysis is related to the H\"older or Hurst exponents, including their respective generalizations. Although different definitions are possible, the conceptual goal is always the same: to describe how the function evolves around a given point by converging to a value (regular) or diverging (singular).

In the most general case, given a signal $s$ that is defined on $\Real^d$ domain and images to a $\Real^m$ space, the H\"older exponent $h(\px)$ of point $\px$ is the exponent satisfying the following limit, when it exists \cite{Jaffard.1997} :

\be
\parallel s(\px+\pr)-s(\px) \parallel\; =\; \alpha(\px) r^{h(\px)} \; +\; o(r^{h(\px)}) \qquad (\pr \rightarrow 0)
\label{eq:Holder}
\ee

\noi
where $r = \| \pr \| $. This means that in the proximity of $\px$ the signal follows a power law of exponent $h(\px)$. An alternative definition that analytically is slightly more restrictive is usually called the Hurst exponent \cite{Simonsen.1998,Jones.1996} and defined as $s(\px+\pr)-s(\px)  = \langle{\bf \alpha(\px)}| \pr \rangle \; r^{h(x)-1} \; +\; o(r^{h(\px)})$ where ${\bf \alpha(\px)}$ is a continuous $(1,1)$ tensor. For the purpose of this article, analysis of 1D signals of 1 component the definitions actually coincide.

The concept of singularity exponent can be interpreted also in terms of differentiability. A function that is strictly $n$-derivable at point $\px$ has a singularity exponent $h(\px)=n$. So that in this sense the singularity exponent can be related to non-integer differentiability. In a similar way, as we will see below, it is also related to the content of information.

Nevertheless, H\"older or Hurst exponents defined this way have very specific applicability (e.g., in the case of multiaffine functions) and cannot be directly found in real-world signals. The main reason is that the basic power-law behavior is masked by the presence of long-range correlations, noisy fluctuations, discretization and finite-size effects. All these make that the analytical limit described is not practically attainable \cite{Turiel.2000a,Turiel.2008}, and a generalized definition of singularity exponent is needed. To achieve this, the objective is to find a certain measure $\mu$ for which we could take a similar limit:

\be
\mu \left({\cal B}_r(\px) \right)\; =\; \alpha(\px) \: r^{d+h(\px)} \: +\: o \left( r^{d+h(\px)} \right) \qquad (r \rightarrow 0)
\label{eq:MF}
\ee

\noindent
where $d$ is the dimension of the domain, i.e., $d=1$ in the 1D case, and ${\cal B}_r(\px)$ is a ball centered around $\px$ having a radius $r$ for a certain norm (choice to be done for multi-dimensional cases; they all coincide in 1D).

The actual definition of singularity exponent that we will be using in the present article works well in practice and is little affected by the artifacts mentioned above. For it, we will work on the gradient-modulus measure of the signal \cite{Turiel.2000a}. This measure is defined from its density:

\be
{\mbox d}\mu (\px)\; =\; \| \nabla s\|(\px)\: {\mbox d}\px
\label{eq:dmeasure}
\ee

\noindent
a definition that is absolutely continuous with respect to the Lebesgue measure. Hence, the measure of any Borelian ${\cal A}$ is given by:

\be
\mu ({\cal A})\; =\; \int_{\cal A} \! {\mbox d}\px \: \|\nabla s\|(\px)
\ee

The gradient-modulus measure characterizes the local singularity of any point. A signal that has a H\"older exponent $h(\px)+1$ according to eq.~(\ref{eq:Holder}) will fulfill also eq.~(\ref{eq:MF}), with this $+1$ shift.

Practical calculations of eq.~(\ref{eq:MF}) can benefit from using wavelet-projected interpolations, this way effectively avoiding some of the discretization effects \cite{Daubechies.1992,Mallat.1999}. The wavelet projection of the measure at point $\px$ and scale $r$ is expressed as ${\cal T}_{\Psi} \mu (\px,r) = \int_{\Real^d} {\mbox d}\mu(\px^{\prime}) \, {r^{-d}} \, \Psi \left((\px -\px^{\prime}) / {r} \right)$  with $\Psi$ being a predetermined function known as the mother wavelet. As we can see, the operator ${\cal T}_{\Psi}$ is a map from the set $ \cal M$ of $\sigma$-finite measures on  $\Real^d$ to the set of functions $\Real^d \times \Real^+ \rightarrow \Real$. That is why a signal that has a singularity exponent at the point $\px$ according to eq.~(\ref{eq:MF}) exhibits this same exponent when wavelet-projected \cite{Daubechies.1992,Turiel.2000a}, i.e,

\be
{\cal T}_{\Psi} \mu (\px,r)\;=\; \alpha_{\Psi}(\px) \: r^{h(\px)} \: +\:
o \left( r^{h(\px)} \right)  \qquad (r \rightarrow 0)
\label{eq:MF-wav}
\ee

It is worth mentioning that wavelet projections expressed in this way treat the wavelet function as a kernel for the measure and no additional restriction is imposed. This way, we are not limited to use only admissible wavelets (i.e., wavelets that form a basis of a function space). In particular, we can use always-positive kernels that do not have zero-crossings. High-order wavelets that exhibit several zero-crossings have a significant loss in spatial resolution \cite{Turiel.2003,Turiel.2006}, but positive kernels minimize spatial spread and can normally reach the original resolution, that is, one pixel in the original signal.

\subsection{Reconstruction formula}

The existence of a multiscale hierarchy details dynamical redundancies in the signals that can be exploited to reconstruct them from partial information \cite{Turiel.2002}. In the following, we describe a reconstruction algorithm for multiscale signals that is model-agnostic and only assumes empirically verifiable hypotheses on them.

The starting point is equation (\ref{eq:MF-wav}): the signals under analysis exhibit a singularity exponent $h(\px)$ at each point \cite{Parisi.1985,Frisch.1995}. In addition, the values of the singularity exponents are organized so that they define a hierarchy of multiscale geometrical structures \cite{Falconer.1990}. In the literature, these two hypotheses are commonly verified statistically at a global level \cite{Falconer.1990,She.1994} but they can be expressed also at each point if using a specially appropriate multiscale functional \cite{Turiel.2002,Turiel.2000a,Turiel.2008}. The reconstruction formula aims at reconstructing the complex signal from the vertex of the singularity exponent hierarchy.

That reconstruction is proved from the theoretical point of view in the case of multifractal functions when the vertex of the hierarchy is associated to the so-called {\bf Most Singular Component} (MSC) \cite{She.1994}, which is the set comprising the points with most singular ({\it i.e.}, most negative) values of $h(\px)$. Therefore it is reasonable to study its reconstruction capabilities on empirical signals, especially when that component constitutes a meaningful attractor of the signal dynamics. This can be achieved in practice by analyzing the information content in the signal to consider the most informative set of points as its MSC. Provided that the reconstruction works, the most informative set actually constitutes an Unpredictable Point Manifold \cite{Pont.2011d}. Signal predictability is a core concept in the processing of complex signals~\cite{Boffetta.2001a,Aurell.1997}, where measures such as Lyapunov exponents and Kolmogorov-Sinai entropy are commonly used to identify information content.

The reconstruction algorithm is designed to act on the gradient of the signal (or the derivative in the case of 1D functions we present here) because the multiscale functional used, eq. (\ref{eq:dmeasure}), is gradient-based and so this point simplifies the development. In order to keep the algorithm as general as possible, we look for a universal operator. This operator reconstructs a signal from its Unpredictable Point Manifold -- ideally, from its MSC \cite{Turiel.2002} -- and it is consistent with the known statistical invariances of multiscale signals \cite{Frisch.1995}.

The algorithm is model-agnostic in the sense that it does not assume any particular distribution or correlation structure for the MSC set of points or the associated singularity exponents. The only assumptions on the projecting operator is that it is required to be deterministic, linear, translational invariant, isotropic and leading to the known power-spectrum shape. For the case of time series, there is an alternative description that substitutes the isotropy condition by the causality condition. These requirements uniquely define the operator \cite{Turiel.2002}. 

For a given multiscale signal $s$ let us denote by $\MSC$ its MSC, i.e., the set of points $\px$ such that $h(\px) \in ] h_{\infty} -\Delta,h_{\infty} +\Delta [$. $h_{\infty}$ is the minimum value of $h(\px)$ over the signal and $\Delta$ is a small, infinitesimal threshold (ideally). The essential gradient of the signal is defined as $\essgrad{\msc} (\px)= \nabla s(\px)\: \delta_{_{\msc}}(\px)$. In that context, $\delta_{_{\msc}}$ means a delta distribution over the $\MSC$ (it uniformly weighs MSC points and vanishes outside of the MSC).

The reconstruction formula introduced in \cite{Turiel.2002} can be expressed as:

\be
s(\px) \; =\; (\vec{g} \cdot \essgrad{\msc}) (\px)
\label{eq:recons}
\ee

\noindent
where $(\, \cdot \,)$ is the convolution dot-product and the vector field $\vec{g}$ is the reconstruction kernel. That kernel can be expressed in Fourier space in a compact way:

\be
\hat{\vec{g}}(\pk)\; =\; i \frac{\pk}{\|{\pk}\|^2}
\ee

\noindent
with $i=\sqrt{-1}$ as the imaginary unit, $\pk$ as the frequency vector and the hat $\hat{ }$ indicating Fourier transform.

Furthermore, the reconstruction formula determines how the MSC is related to the most informative set of points in the signal. The derivation in \cite{Turiel.2002} implies that any set ${\cal F}$ reconstructing the signal must verify:

\be
\mbox{div} \:\left(\essgrad{{\cal F}^c}\right)\; =\;0
\label{eq:UPMbasis}
\ee

\noindent
where ${\cal F}^c$ is the complementary set of ${\cal F}$. That applies in particular to the MSC, but since the divergence operator is local and eq. (\ref{eq:UPMbasis}) is linear, the reconstructibility of a point can be decided based on its neighborhood only. Therefore, the MSC points are those that 
must be included to ${\cal F}$ to have a full reconstruction of the signal, because their values cannot be predicted just knowing the values in their surroundings. In other words, the MSC points are the unpredictable points of the signal \cite{Turiel.2008}.

\section{Atrial fibrillation}
\label{sec:data}

Atrial fibrillation (AF), the most common form of cardiac arrhythmia, is responsible for significant morbidity each year in all parts of the world. It results from the chaotic operation of the top of the heart (atria). Although a priori it is not a severe condition by itself, it causes a high mortality rate by its most severe complications, either from heart failure or by stroke-related embolism. The main cause of AF is related to a change in the electrical conduction properties in some type of cardiac tissue: some areas of this tissue depolarize spontaneously or slow the spread of the pulse. Today the primary treatment for AF remains medication, but ineffective drugs, intolerance to them or their side effects has led to the development of new forms of treatment, mainly radiofrequency ablation. This ablation technique requires the introduction of a catheter within the heart in order to burn the areas that have become electrically deficient. A low-voltage radiofrequency current is applied to the pathogenic areas heating them to induce the necrosis of the tissue at tiny width and depth (in the order of few mm). In cases of paroxysmal AF, Ha\"issaguerre et al. have shown \cite{Haissaguerre.1998} that for 80\% of patients, the pathogenic tissue is located in one of the 4 pulmonary veins, and electrical insulation, obtained by surgical means, allows the patient to regain a normal heart rhythm \cite{Dubois.2009,Sanders.2006,Takahashi.2006,Hocini.2005,Hocini.2005a,Hocini.2005b,Haissaguerre.2003}. But in more severe cases, i.e., persistent or permanent AF, locating the pathogenic areas remains difficult and is still an unsolved problem. During the ablation procedure, catheters inserted in the heart can analyze finely the electrical activity of the atria. The morphology of the signals obtained and their temporal evolution must guide the surgeon to the location of the sources to ablate. But the complexity of the acquired signal makes analysis very difficult and there is no clear identification of important information. It is in this context where nonlinear analysis techniques like the MMF can be applied to identify the changes in cardiac regime that lead to the recovery of the normal sinus rhythm.

\subsection{The data}

\begin{figure}[hbtp]
\begin{center}
\includegraphics[width=0.49\textwidth]{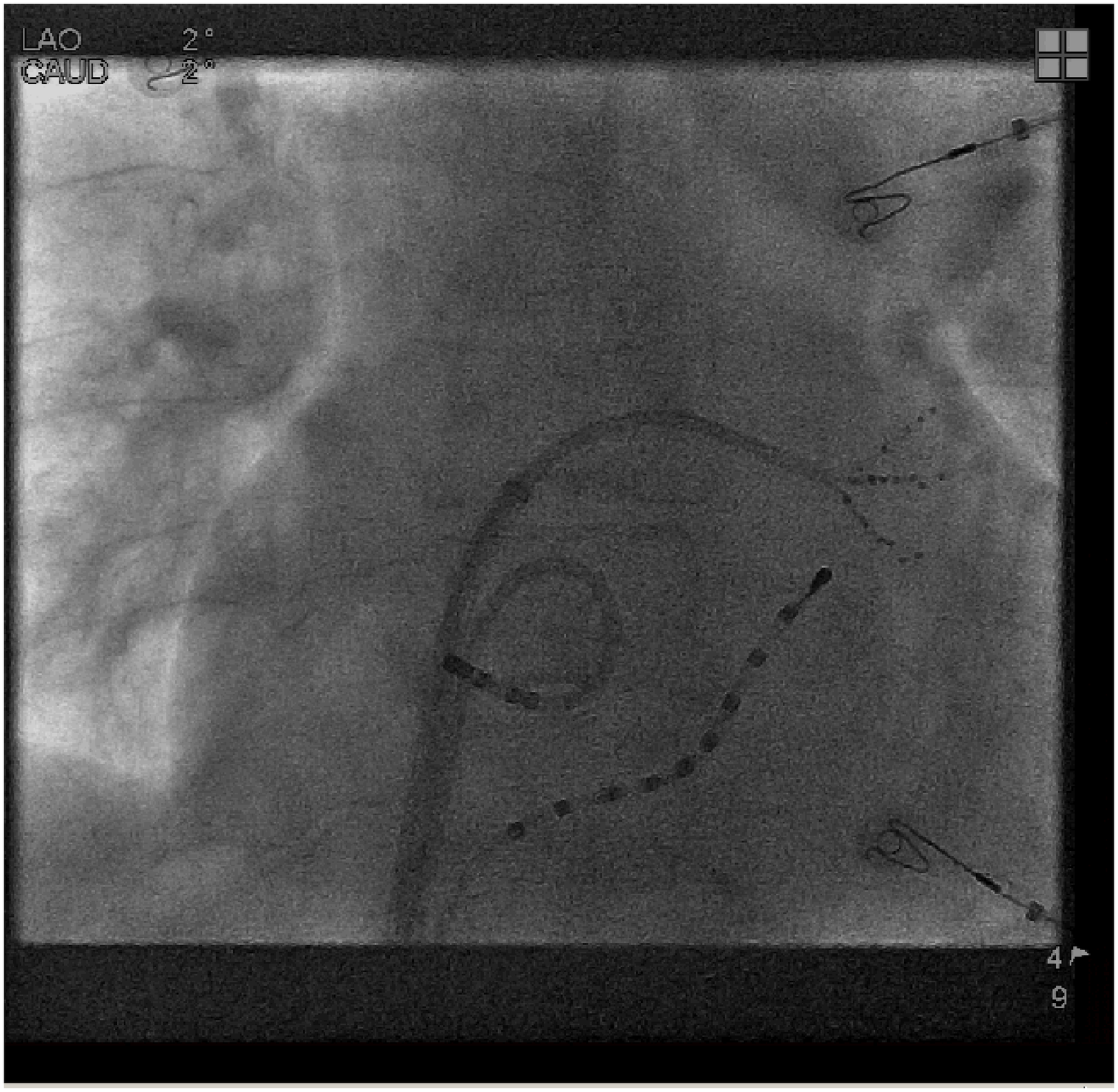}
\includegraphics[width=0.49\textwidth]{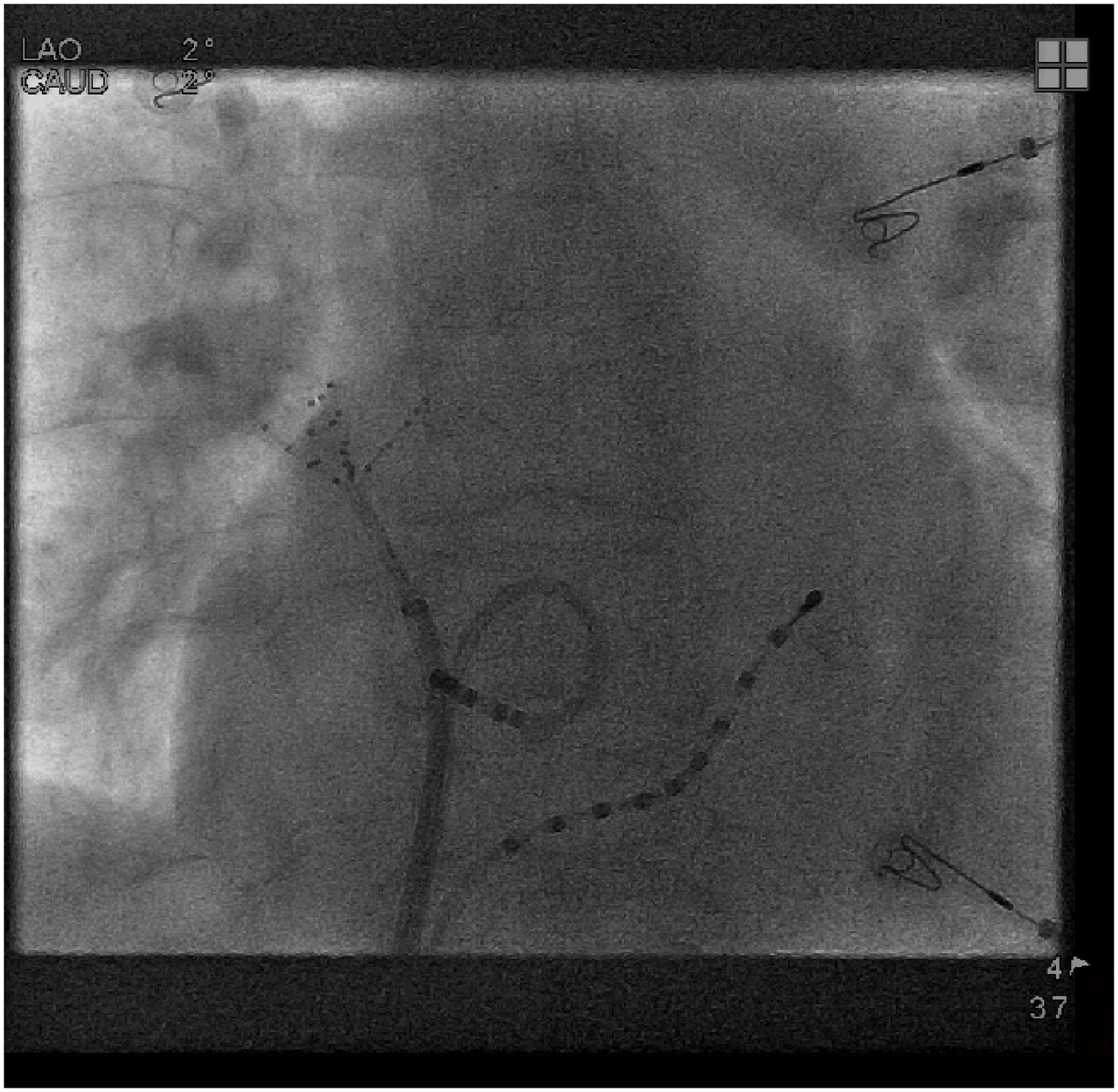}
\end{center}
\caption{Radiographic images taken during the radiofrequency ablation procedure and showing the three catheters and the electrode placement. In both cases, we can observe an electrode catheter measuring along inside the coronary sinus, the ablation catheter and the pentaray catheter taking measures. The ablation catheter is placed curved and was not operating at the moment of the radiographies (it is nevertheless always taking measures also). The pentaray catheter is shown taking measures in the left inferior pulmonary vein area (left image) and in that of the right superior pulmonary vein (right image).}
\label{fig:radio}
\end{figure}

The processed data consist of multilead recordings of the electric potential measured on six human patients affected by different types of atrial fibrillation who have undergone radiofrequency ablation in their atrial endocardia. Measures have been taken before, during and after RF ablation at the Haut-Leveque hospital in Pessac. There are both unipolar and bipolar leads, and there are ECG measurements on the skin concurrently with electrode catheters inside the heart. The standard ECG leads recorded are I, II and III (bipolar) and V1 (unipolar); these constitute the first 4 data channels in our datafiles. Three electrode catheters are introduced through the main veins of the patient. One of them is the radiofrequency catheter, which is used for the ablation but is able to measure the potential as well; it consists of four electrodes and so provides potential measures at two positions, namely the distal one and the proximal one. The second one is called pentaray and it is a catheter that can be opened at its end dividing in five branches that spread equiangularly; each branch has four electrodes measuring at a distal and a proximal positions. These two catheters are introduced inside the heart. A third catheter containing ten electrodes is introduced to the coronary sinus and provides measures at five positions. In total we have 4 ECG potential measures on the skin and 17 intracardial measures for each time instant. Signals are sampled at a rate of 1 kHz for a total of 1 183 232 data points. Figure~\ref{fig:radio} shows radiographies illustrating the electrode placements.

\begin{figure}[hbtp]
\begin{center}
\includegraphics[width=0.49\textwidth]{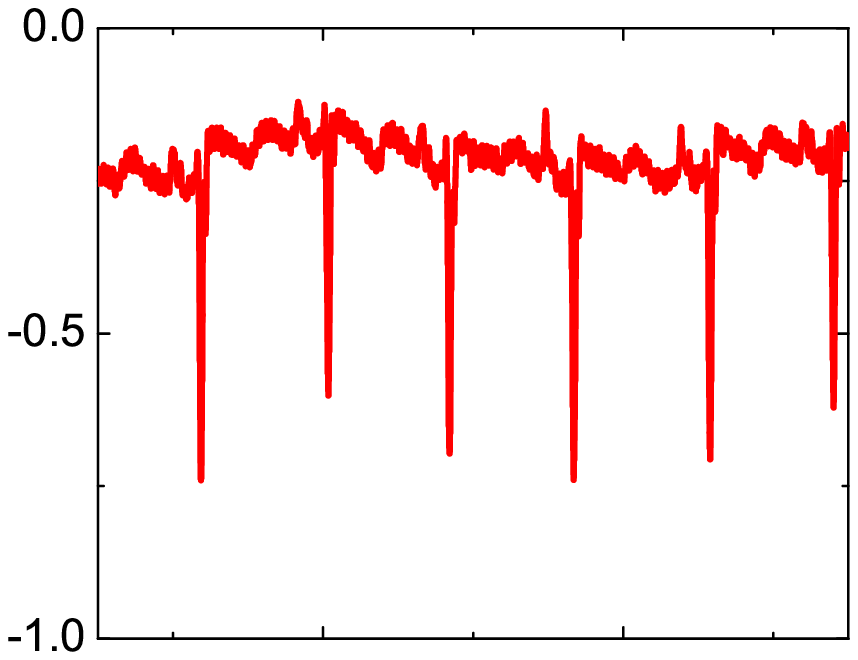}
\includegraphics[width=0.49\textwidth]{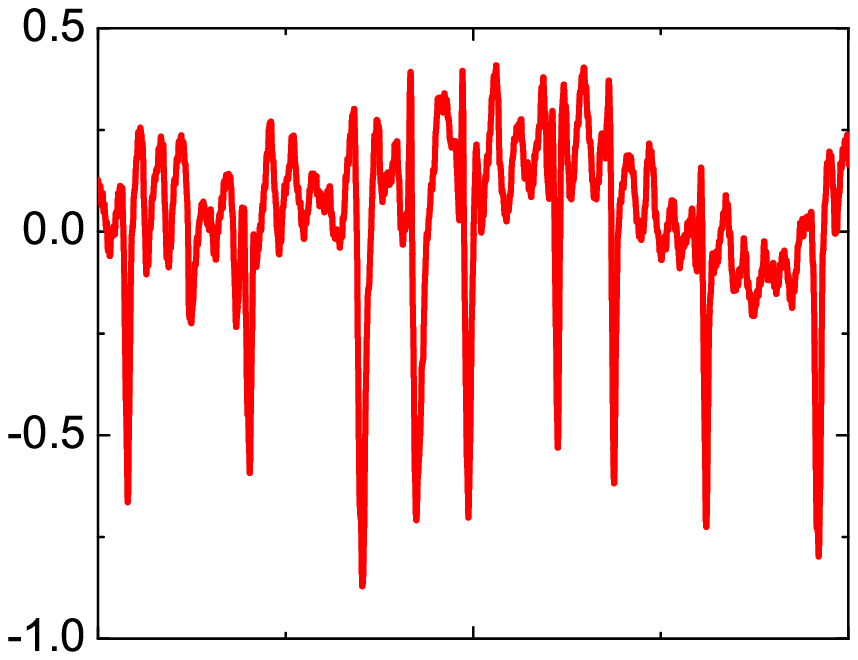}
\\
\includegraphics[width=0.49\textwidth]{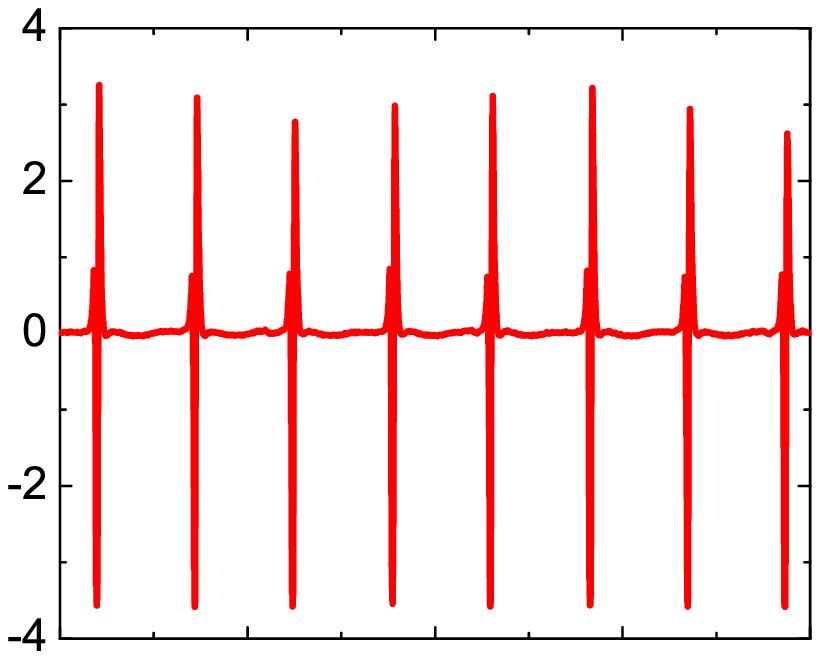}
\includegraphics[width=0.49\textwidth]{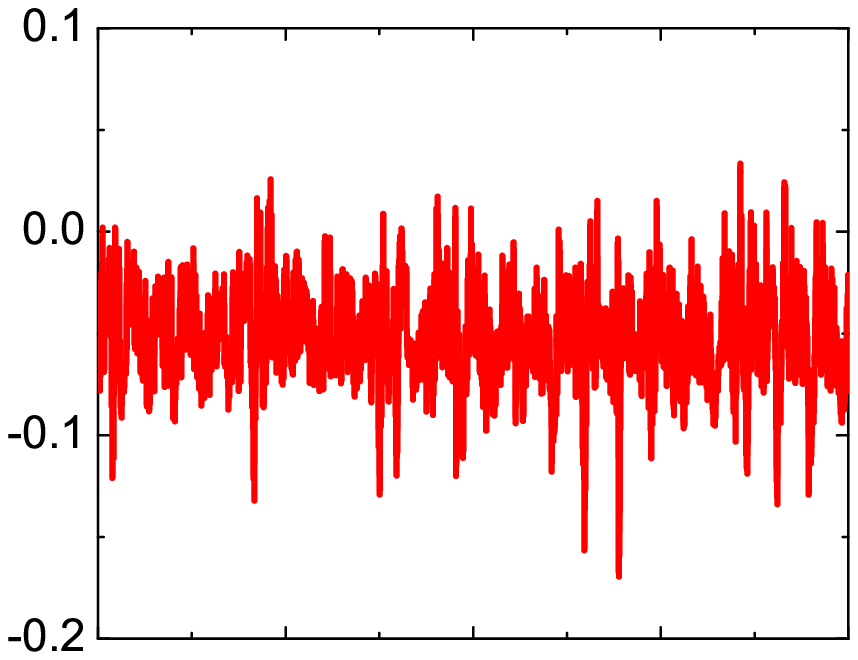}
\end{center}
\caption{Illustration of normal and atrial fibrillation heartbeats. At the top, we show external electrocardiogram measures, namely from V1 electrode, for the case of sinus rhythm (left) and atrial fibrillation (right). At the bottom, we show the measurements of the first pair of electrodes of the catheter on the endocardial area around the right superior pulmonary vein. Fibrillation is clearly seen inside the heart, and can be still noticed outside.}
\label{fig:af}
\end{figure}

For illustration purposes, we show in Figure~\ref{fig:af} how the fibrillation (desynchronized beat) is seen in the right atrium and how AF is still slightly manifested in the measure from the electrode on the skin V1.

As a complement, we have also processed data from MIT-BIH Arrhythmia Database \cite{Moody.2001,Goldberger.2000}. These consist of long and extensive ECG measurements under different arrhythmic regimes, in many cases containing also the transition period between regimes. The most remarkable particularity is probably the fact that they are annotated at the beat level by multiple experts independently (and these annotations have been audited and verified by the full community of the database users since 1980). They consist of two channels: a bipolar lead (II) and a unipolar V$^\ast$ lead, sampled at 360 Hz for 30 min. for a total of 47 patients of Boston's Beth Israel Hospital between 1975 and 1979. Electric signals were originally recorded analogically into magnetic tapes, but are today distributed as digital files. We have processed files 200 to 205 as those appeared to be the ones with most AF episodes.

The significant differences between one dataset and the other made us expect important differences in their analyses, particularly in case spurious effects had appeared. On the contrary, we have observed consistent results, which indicates the robustness and meaningfulness of the analyzing methodology performed.


\section{Results: Heartbeat analysis}
\label{sec:results}

\begin{figure}[hbtp]
\begin{center}
\includegraphics[width=0.8\textwidth]{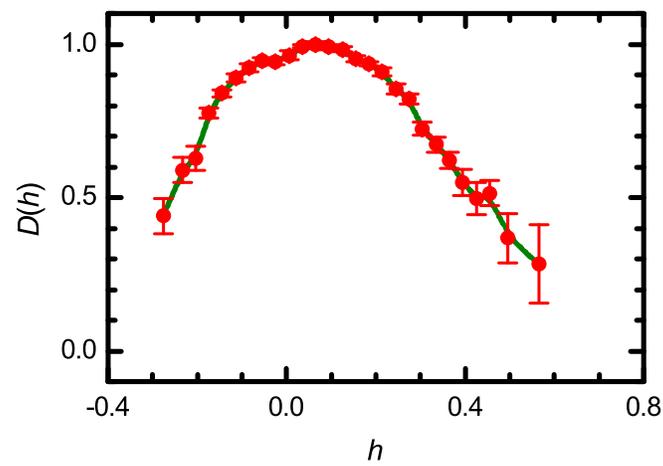}
\caption{Singularity spectrum showing the Hausdorff dimension of exponent level sets in front of singularity exponent values for the processed heartbeat data. As expressed in Section \ref{sec:MMF}, the computed singularity exponents refer to the gradient-modulus measure and so singular (negative) values do not imply discontinuities in the signal but in its gradient, as there are no exponents below -1. In addition, the full-support component, i.e. that of dimension 1, corresponds to a positive singularity value, thus ensuring the integrability of the trajectory.}
\label{fig:spectrum}
\end{center}
\end{figure}

The first step to perform on the empirical signals consists on the validation of the multifractality hypothesis under the MMF. Previous studies have validated multifractal dynamics under a canonical formalism, which is based on statistical averages (the scaling of structure functions or partition functions). However, the MMF is directly geometrical instead of statistical, which means that we characterize the multiscale character at each point of the signal. In Figure \ref{fig:spectrum} we show the resulting singularity spectrum estimated for the ensemble of signals \cite{Pont.2009,Pont.2011b}. Convergence to the presented curve has been observed for all the individual signals between the empirical error bars. 

Once the heartbeat data has been validated to be microcanonical, i.e., it fits the singularity-exponent relation seen in eq. (\ref{eq:MF-wav}), the next step consists in analyzing the dynamical properties and characterizing transition points. A first observation is that the measures inside the heart clearly have more abundance of most singular points. This is something that we expected is because the difference observed -- more singular points -- means that the internal signals are more informative and require more points to be reconstructed. Multifractal signals can be reconstructed from their component of most singular points (which therefore concentrates all the information), if such component exists \cite{She.1994}. In our case, the most singular component (i.e., the most informative) is also larger and more dispersed in the internal than in the external measures. We show below that the studied signals are effectively reconstructible though nevertheless, the key dynamical parameters detected in internal measures are still noticeable in the external measures.

A general theory and methods about reconstructibility of multifractal signals from the Most Singular Component (MSC) can be found in \cite{Turiel.2008}. With the MMF we have access to the singularity value of each point, which is the basis of the multiscale coordination of the components in the signal. Additionally, the methodological tools provided by this formalism are effectively adapted to work with real signals (with discretization, noise, artifacts, aliasing, correlations), and we can use this information to look at the dynamical properties of the heartbeat series.

\subsection{A dynamical model for analysis}

Given the points we have, we propose a dynamical model for the analysis of heartbeat signals. The first key element comes from the calculation of the singularity exponents and, in particular, their associated Most Singular Component (MSC). An interesting aspect of the MSC is that we do not need the values of the signal over its points nor the actual values of singularity exponent. In fact, most of the dynamical information of the signal is contained only in the orientation over the MSC \cite{Turiel.2003b} and we can use an analysis of this orientation as the starting point.

First of all, we calculate the singularity exponents of the signals to determine where their MSC is located. To achieve this, we locate where are the smallest (most singular) exponents in the series, giving a small tolerance level to account the numerical fluctuations. Then we define the oriented MSC as a function that is zero everywhere except for the most singular points, where it takes the value of the sign of the gradient.

When we apply the reconstruction formula defined in \cite{Turiel.2002} to the oriented MSC, the result is a {\it reduced} signal that coincides with the original signal at short scales. At long scales, there is a slow divergence between them. This allows to model the dynamics as a combination of a fast dynamics driven by the MSC orientation and a slow-varying field that acts as a factor on it \cite{Turiel.2005}.

More concretely, we define the oriented MSC as $\delta_\infty(t)$ taking +1 on MSC points of positive derivative, -1 on MSC points of negative derivative and 0 on non-MSC points. This way, transitions from one point to the other can be described as a Markov chain. This way, we call $\sigma$ the Markov states. The two-point joint probability is noted as:

\be
P(\sigma_0,\sigma_\tau)=\left\langle P(\delta_\infty(t)=\sigma_0, \delta_\infty(t+\tau)=\sigma_\tau) \right\rangle
\ee

\noi
and as a consequence the marginal probabilities are:

\be
P(\sigma_0)=\left\langle P(\delta_\infty(t)=\sigma_0) \right\rangle_t = \left\langle P(\sigma_0,\sigma_\tau) \right\rangle_{\sigma_\tau}
\ee

Under the hypothesis of distributional stationarity, we can expand the process as transitions between all the states \cite{Turiel.2003b}. The transition at two steps is expressed as:

\be
P(\sigma_2|\sigma_0) = \sum_{\sigma_1} P(\sigma_2|\sigma_1) P(\sigma_1|\sigma_0)
\label{eq:two-steps}
\ee

\noi
i.e., twice the one-step transition. This means that we can represent all the Markovian dynamics through $P(\sigma_1|\sigma_0) $. The matricial expression of the process gives the so-called transition matrix. In our case, with three possible states: +1, -1, 0.

\be
T=
\left( {\begin{array}{ccc}
 00 & 0+ & 0- \\
 +0 & ++ & +- \\
 -0 & -+ & -- \\
 \end{array} } \right)
\ee

In a Markov process, a state is called {\it recurrent} when the probability of starting at that state and returning to it after a finite number of steps is one. A stationary distribution for the process exists if and only if all the states are recurrent and all the expected times of first return are finite. This is always the case in our process, because all transitions are possible and none is absorbing. That stationary marginal distribution is a state of convergence of the process. Applying $T$ to it results the same, so this is the first eigenvector of the process with an eigenvalue of 1. The other two secondary eigenvalues give a characterization of the dynamics.

\subsubsection{Hands on the data:}
The first calculation to be done is the validation of eq. (\ref{eq:two-steps}): for all the signals the differences between the empirical two-steps $P(\sigma_2|\sigma_0)$ couple values and those derived from combination of two  $P(\sigma_1|\sigma_0)$ steps have a median relative discrepancy of 1\%. These differences are compatible with the empirical estimation sampling errorbar and mean that in all cases the memory of the MSC orientation decays very fast so that the Markovian behavior hypothesis for it could be reasonably assumed.

A preliminary look at the oriented MSC distributions does not show sharp differences between a signal and another. The estimation of the eigenvalues of $T$ matrices are primarily limited by their least probable elements. Sampling error propagation is around 7\% for each lead and heartbeat regime in the Haut-Leveque database and 3\% for each lead and heartbeat regime in the MIT-BIH arrhythmia database.

To this effect, we have analyzed the Markov processes for all the data, classified in four categories: internal channels under Atrial Fibrillation (AF) internal channels under (normal) sinus rhythm, ECG channels under AF and ECG channels under sinus rhythm. We did not observe significant differences from one patient to the other or from one channel to the other inside the category, so we have grouped them to enhance presentation of the results and maximize the precision. That said, when processing one single channel the results are already stable, so we conclude that the method is robust and little data demanding.

\begin{figure}[hbtp]
\begin{center}
\includegraphics[width=0.6\textwidth]{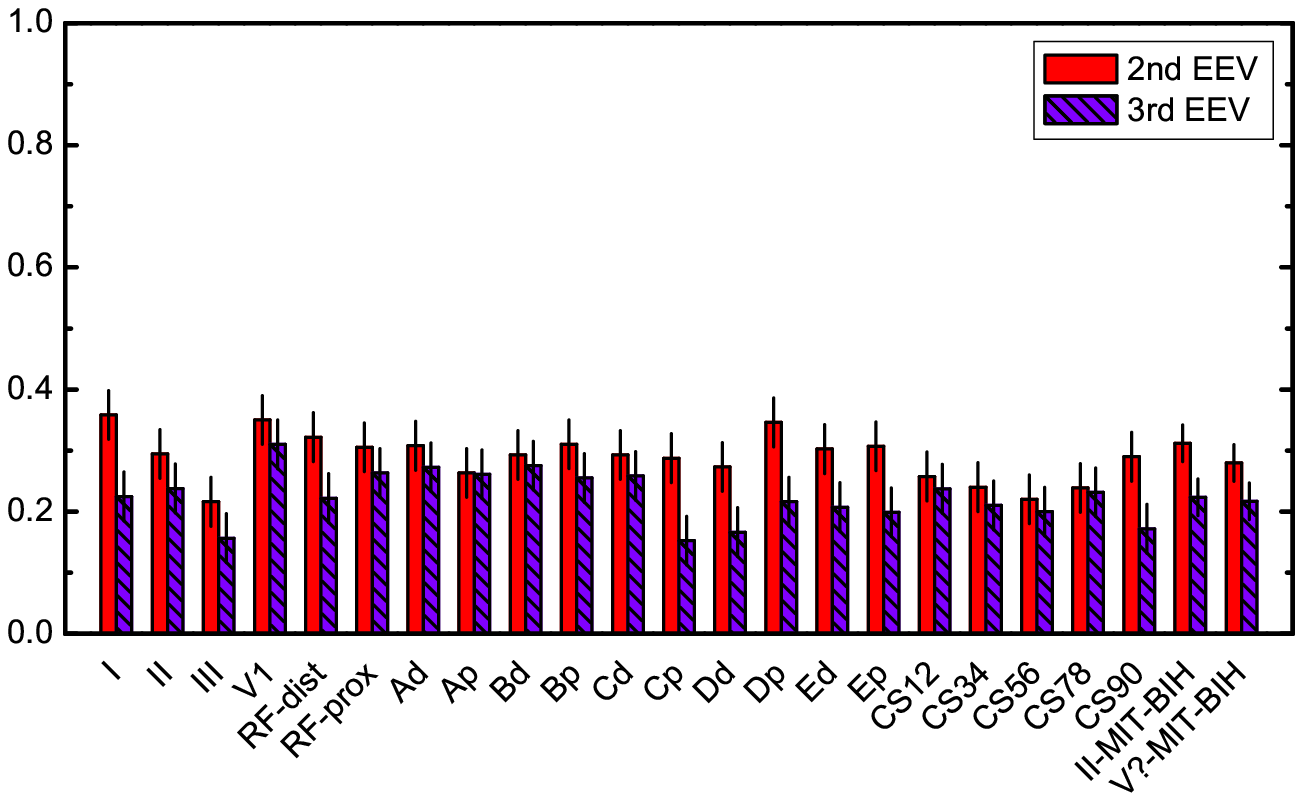}
\includegraphics[width=0.6\textwidth]{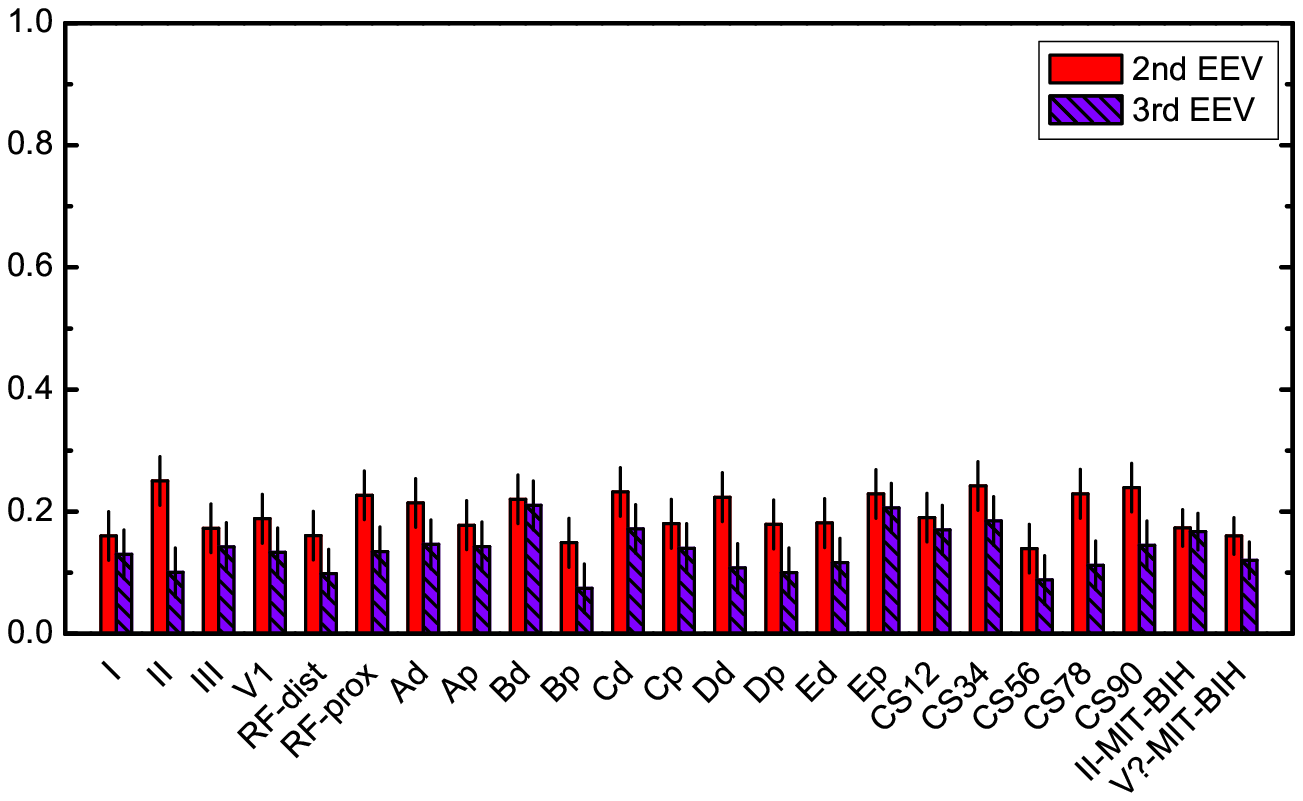}
\includegraphics[width=0.6\textwidth]{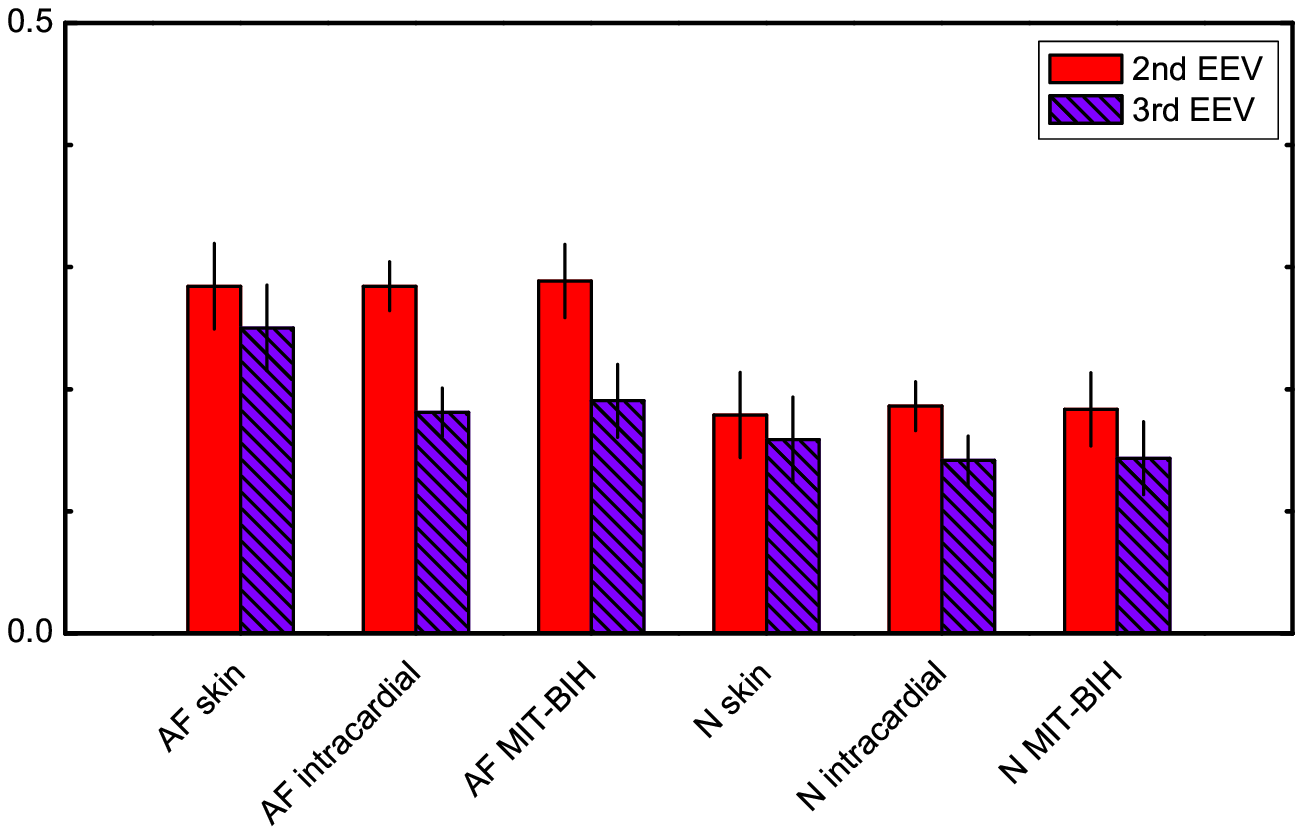}
\caption{Transition matrices $T$ of the oriented MSC have been calculated for each signal in each regime; we show their respective second and third eigenvalues which are the parameters leading that orientation dynamics for the atrial fibrillation cases (top panel) and the normal sinus rhythm (middle panel). Correspondences between channels show that we can group all the internal and all the external leads for each regime thus further reducing the sampling error; that is shown in the bottom panel with 6 signal groups: Haut-Leveque internal (intracardial) measures, Haut-Leveque ECG measures and MIT signals, each for healthy and AF cases. Numerical values are given in Table~\ref{tbl:results}}
\label{fig:results}
\end{center}
\end{figure}

As we can see in Figure \ref{fig:results}, the particular signature of AF is conserved when the signal inside the heart is propagated to the skin. We notice that despite the signals clearly differ in appearance, the dynamical parameters in terms of MSC orientation dynamics for the internal measures are still externally observed with minimal differences, while at the same time the AF condition significantly perturbs these values both internally and externally. The same behavior is confirmed at processing the MIT-BIH arrhythmia database signals. This suggests that it should be possible to finely monitor the AF evolution and severity from external ECG measures by means of advanced statistical measures that are stable and robust. Even more, transitions to and from fibrillation could be immediately detected.

\begin{table}[hbtp]
\begin{center}
\begin{tabular}{| c | c c c c |}
\hline
~~Signal group~~  &  ~~Second eigenvalue~~  &  ~~Second-eigenvalue error~~  &  ~~Third eigenvalue~~  &  ~~Third-eigenvalue error~~\\
\hline
AF skin		&	0.28 &	0.04 &	0.25 &	0.04\\
AF intracardial &	0.28 &	0.02 &	0.18 &	0.02\\
AF MIT-BIH	&	0.29 &	0.03 &	0.19 &	0.03\\
N skin		&	0.19 &	0.04 &	0.16 &	0.04\\
N intracardial	&	0.19 &	0.02 &	0.14 &	0.02\\
N MIT-BIH	&	0.18 &	0.03 &	0.14 &	0.03\\
\hline
\end{tabular}
\caption{Values for 2nd and 3rd eigenvalues and their respective errors as plotted in Figure~\ref{fig:results}}
\label{tbl:results}
\end{center}
\end{table}

\subsubsection{Source field:}

As we have mentioned above, the oriented MSC drives the fast dynamics, meaning that the sign clustering determines a big part of the dynamical structure. Nevertheless, on a longer scale the MSC alone does not precisely reconstruct the signal because the {\it constant} factor is not really constant but slowly evolves. This leads to the definition of the source field. Given a signal $s$ and the reduced $r$ constructed from its oriented MSC, the source field $\rho$ is defined such that:

\be
\nabla s (x) = \rho (x) \, \nabla r (x)
\ee

\noi
but nevertheless this definition is not usable in practice in points outside the MSC, because the reduced signal is not well defined in them. To solve this, we make use of a more rigorous definition of source field in terms of measures \cite{Turiel.2005}:

\be
\mu_s({\cal A}) = \int_{\cal A} {\rm d}\mu_r(x) \rho(x)
\label{eq:sources}
\ee

\noi
which implies that the source field $\rho(x)$ is the Radon-Nikodym derivative of the two measures: $\rho(x) = {\rm d}\mu_s / {\rm d}\mu_r$.

\begin{figure}[hbtp]
\begin{center}
\includegraphics[width=0.49\textwidth]{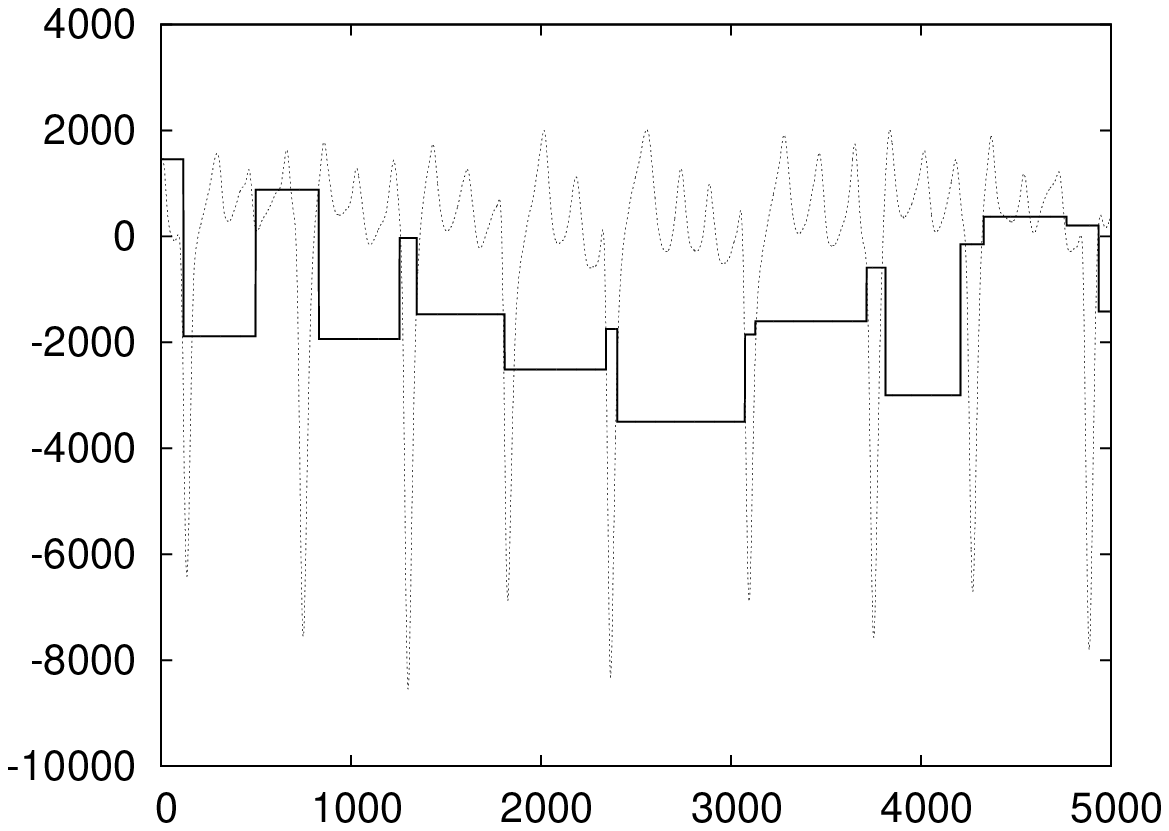}
\includegraphics[width=0.49\textwidth]{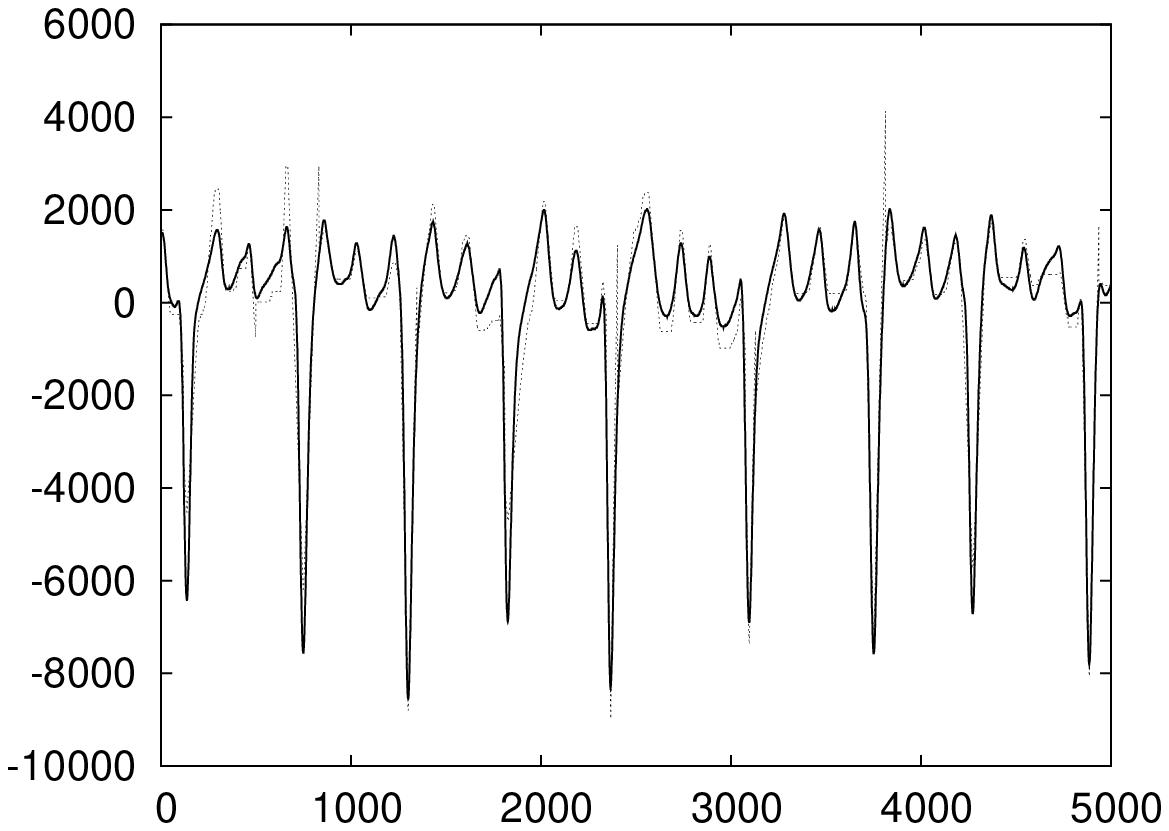}
\end{center}
\caption{Time evolution of the source fields and the reconstructions of an illustration signal corresponding to measures from V1 electrode under atrial fibrillation. At the left, source field displayed (solid) over the original signal (dashed). At the right, signal reconstruction (dashed) based on the source field and the Markov-chain modeled MSC. Signs of Atrial Fibrillation are noticed in the dynamical parameters. Reconstruction is of pretty good quality, especially for the peaks.}
\label{fig:sources}
\end{figure}

There exist several different strategies to numerically estimate a Radon-Nikodym derivative. Since in our case we are most interested in the detection of slow dynamical transitions, we have used an iterative algorithm that fits eq. (\ref{eq:sources}) in a piecewise constant fashion. This way, we concentrate on the determination of the dynamical borders and let the MSC lead all the fast Markovianly-stable evolution. The results, shown in Figure \ref{fig:sources} show that the source field varies infrequently though it exhibits quite sharp transitions. In consonance with the results for the MSC orientation in the subsection above, the dynamical character in the case of AF is significantly different than under sinus rhythm.

We observe a correspondence of the transition points with the points in which the deviation between the original and the reconstructed series is more important. So these transition points correspond to transitions in the reconstructibility and in the content of information, which means that the detected transitions correspond to actual changes in the dynamical properties of the signal. The concrete mechanism that establishes the link of correspondence with some electrophysiological transition is nontrivial and complex.


\section{Conclusions}
\label{sec:conclusions}

We have performed a detailed study of cardiac fluctuation signals under the approach of the Microcanonical Multiscale Formalism (MMF). The structural complexity inherently present in those signals challenges their analysis. The emergence of a complex dynamic behavior hinders most attempts to carefully estimate microscopic dynamical parameters in the system, especially unless cardiac models are assumed with heavy calibration or data assimilation. Therefore, an analysis describing in terms of effective dynamics is especially appropriate for heartbeat. These signals have been known to be of multiscale character since several decades ago \cite{Kitney.1980,Kobayashi.1982,Peng.1993} but surprisingly enough this aspect is presented most of the times in the literature as coincidences between separate scale levels, without exploiting the capabilities of multiscale structure and mutual relationships -- of course with notable exceptions such as in \cite{Ivanov.1999}. The description in terms of multiscale effective dynamics is essentially model-free, and it assumes hypotheses on the data that are easily verifiable. In that sense, this approach is close to that of Machine Learning methodologies, but has the advantage that it tracks closely the key exponents of the emergence of complexity in the system and so the parameters with a physical meaning, be it in terms of information content, information transfer, dynamical attractors or critical sets \cite{Pont.2009}.

The multiscale structure in heartbeat is reflected as a definite geometrical distribution around manifolds of singularity. This way, the signal under analysis is decomposed into different components depending on their characteristic singularity exponent. This fact decomposes the signal into separate regimes according to their characteristic dynamics. The value of the singularity exponent characterizes the power-law behavior under scale changes and directly indicates the information content of the component. Consequently, our analysis provides direct access to the dynamical structure at each point of the signal and it unveils the geometry of singularity components and characterizes the degree of information contained in them. When further exploited, this analysis shows that the most singular component (MSC) contains the information of the entire signal and can restore it. In other words, this component drives the dynamics of the signal.

A first observation is that we reproduce under the MMF the same type of multifractality characterizations for heartbeat series that have been reported under a canonical framework \cite{Ivanov.1999}. Moreover, the expected observation that data from catheters inside the heart have a richer, more informative, more singular multiscale signature than those taken on the skin, does not seem to affect the MSC parameters of the signal or its reconstructibility, which means that the key dynamical features can still be detected on the skin.

In addition, we have shown that the characteristic dynamical parameters retrieved from MMF analysis --namely the orientation of the MSC-- can be dynamically described as a Markov chain. Furthermore, these dynamic parameters (in particular, the transition eigenvalues) under atrial fibrillation (AF) are significantly different from the sinus-rhythm case. Additionally, they can be equally detected from intracardial electrodes or from standard electrocardiogram measures on the skin. Therefore, a possible application would be early detection of transitions to or from the AF. The main database processed consists of potential measures taken during AF ablation procedures at Haut-Leveque Hospital, and the robustness of the approach is confirmed by similar results obtained for the MIT-BIH Arrhythmia Database which contains significantly different signals.

Finally, we have noticed that the signal reconstructibility implies that it can be separated into the fast MSC-orientation process and a complex slowly-varying source field that modulates it. These source fields accurately describe the multifractal dynamic changes, what would suggests a possible relationship with transitions in electrophysiological processes, such as the evolution of the cardiac regulatory mechanism and changes in conductivity of the tissue, as well as other structural changes and drifts.

%
%
\bibliography{biblio}
\bibliographystyle{splncs03}

\end{document}